\documentclass[preprint]{revtex4-1}
\usepackage{multirow}
\usepackage{graphicx}
\usepackage{bm}
\usepackage{dcolumn}
\usepackage{amsmath}
\def \lleq {\lower0.9ex\hbox{ $\buildrel < \over \sim$} ~}
\def \ggeq {\lower0.9ex\hbox{ $\buildrel > \over \sim$} ~}

\def \beq  {\begin{equation}}
\def \eeq  {\end{equation}}
\def \ber  {\begin{eqnarray}}
\def \eer  {\end{eqnarray}}

\begin{document}
\newcommand{\newc}{\newcommand}

\newc{\be}{\begin{equation}}
\newc{\ee}{\end{equation}}
\newc{\ba}{\begin{eqnarray}}
\newc{\ea}{\end{eqnarray}}
\newc{\bea}{\begin{eqnarray*}}
\newc{\eea}{\end{eqnarray*}}
\newc{\D}{\partial}
\newc{\ie}{{\it i.e.} }
\newc{\eg}{{\it e.g.} }
\newc{\etc}{{\it etc.} }
\newc{\etal}{{\it et al.}}
\newcommand{\nn}{\nonumber}
\newc{\ra}{\rightarrow}
\newc{\lra}{\leftrightarrow}
\newc{\lsim}{\buildrel{<}\over{\sim}}
\newc{\gsim}{\buildrel{>}\over{\sim}}
\title{Constraints of the equation of state of dark energy from current and future observational data by piecewise parametrizations}
\author{Qiping Su$^{1}$}
\email{sqp@hznu.edu.cn}
\author{Xi He$^{1}$}
\email{hexi@hznu.edu.cn}
\author{Rong-Gen Cai$^{2}$}
\email{cairg@itp.ac.cn}
\affiliation{
$^{1}$  Department of Physics, Hangzhou Normal University, Hangzhou, 310036, China\\
    $^{2}$ State Key Laboratory of Theoretical Physics, Institute of
Theoretical Physics, Chinese Academy of Sciences, Post Office Box 2735,
Beijing 100190, China   }

\date{\today}

\begin{abstract}
The model-independent piecewise parametrizations (0-spline,
linear-spline and cubic-spline) are used to estimate
constraints of equation of state of dark energy ($w_{de}$) from
current observational data ( including SNIa, BAO and Hubble parameter )
and the simulated future data. A combination of fitting results of
$w_{de}$ from these three spline methods reveal essential properties
of real equation of state $w_{de}$. It is shown that $w_{de}$ beyond
redshift $z\sim0.5$ is poorly constrained from current data, and
the mock future $\sim2300$ supernovae data give poor constraints of $w_{de}$
beyond $z\sim1$. The fitting results also indicate that there
might exist a rapid transition of $w_{de}$ around $z\sim0.5$. The
difference between three spline methods in reconstructing and constraining $w_{de}$ has also been discussed.
\end{abstract}

\maketitle
\section{Introduction}
The current expansion of the universe is  found to be
accelerating~\cite{Riess:1998cb,perlmutter99}, one of the possible
explanations for this is the existence of dark energy(DE), whose
energy density is dominant in the universe and its present equation
of state $w_{de0}$ is less than $-1/3$. At present, it is still fair
to say that one knows a little about the nature of DE. From current
astronomical observational data, we can obtain some properties of
the equation of state $w_{de}$ (the ratio of the pressure and energy
density of DE). To fit the observational data, one has to first
assume a form of $w_{de}$. Some forms of $w_{de}$ have been used in
the literature. For example, the CPL parametrization
$w_{de}(z)=w_0+w_az/(1+z)$\cite{Chevallier:2000qy,Linder:2002et} and
the redshift expansion,
$w_{de}(z)=w_0+w_zz$~\cite{Huterer:2000mj,b1,b2,b3}. It has been
found that $w_{de}$ is very close to $-1$ and should be varying very
slowly with redshift (if any). In the most cases, the cosmological
constant with $w=-1$ is still favored within $2\sigma$ confidence
level (C.L.). Of course the fitting results are dependent on the
parametrization forms adopted. Usually a parametrization form is
only suitable to mimic one type of $w_{de}(z)$. For example, the CPL
parametrization can describe linear and smooth $w_{de}$ well but is
hard to reconstruct $w_{de}$ with oscillations or rapid transitions.
On the other hand, several model-independent methods have also been
proposed to reconstruct
$w_{de}(z)$~\cite{Huterer:2002hy,Huterer:2004ch,Cai:2009ht,Wang:2009sn,Wang:2005yaa},
e.g., the uncorrelated band-power estimate (i.e.,
0-spline)~\cite{Sullivan:2007pd}, cubic-spline
method~\cite{Serra:2009yp}, linear-spline
method~\cite{Cai:2010qp,d1,d2}, wavelet
approach~\cite{Hojjati:2009ab}, and Gaussian process
modeling~\cite{Holsclaw:2010sk,c1,c2}, etc. Most of
model-independent methods have a piecewise $w_{de}(z)$: \be
w_{de}(z)= w_i(z), \ \ z_i<z\leq z_{i+1}, \ee where $w_i(z)$ is a
simple function of redshift $z$, e.g., $w_i(z)$ is just a constant
in each bin for the 0-spline method.

Different model-independent methods should give different but
consistent fitting results of DE. In this paper, we would  like to
get constraints of $w_{de}$ from present and next generational
observations by using three piecewise parametrizations: 0-spline,
linear-spline and cubic-spline. The difference among the three
spline methods in constraining $w_{de}$ will be analyzed. It is
shown that each spline method is only suited to certain types of
$w_{de}$. A combination of constraints of $w_{de}$ from the three
spline methods should help to get real properties of $w_{de}$.

At first, the constraints of $w_{de}$ from current observational
data will be obtained. It is found that the present constraints on
$w_{de}$ are very weak beyond $z\sim0.5$, because in the higher
redshift region there are less data points and the effect of DE on
the expansion of the universe is weaker. Moreover, we see from the
fitting results from 0-spline and linear-spline methods that there
might exist a rapid transition of $w_{de}$ around $z=0.5$. To
estimate the constraints on $w_{de}$ from next generation
observations, $\sim2300$ SN data as forecasted for a space mission
like SNAP/JDEM~\cite{Kim:2003mq,sn300} are simulated. The mock data
are simulated from two fiducial models: one with a smooth $w_{de}$
and the other with a rapid transition $w_{de}$ around $z=0.5$.
In this case $w_{de}$ is poorly constrained after $z\sim 1$.

\section{The constraints on $w_{de}$ from present data}

We  use three spline methods to get constraints of $w_{de}$ from
present observational data, which include type Ia supernovae data
(SNIa), baryon acoustic oscillation (BAO) and observational Hubble
data. The best-fitting parameters and their errors will be obtained
by using the Markov Chain Monte Carlo (MCMC) method. In general, the errors
of $w_{de}$ of different bins are correlated. We will adopt the
decorrelated method proposed in~\cite{Huterer:2004ch} to obtain
errors of a new parameter Q(z). The new parameter $Q(z)$ is defined
by transforming the covariance matrix of $w_{de}$, so that the
errors of Q(z) are uncorrelated and do not entangle in each bin.

\subsection{Observational data}

Since it is hard to get strong constraints of $w_{de}$ in high
redshift region, we will focus on constraining $w_{de}$ in the
region $z\in[0,0.9]$ only in this work. The data sets we adopt are
SNIa Union2 data~\cite{Amanullah:2010vv}, BAO distance parameter
$A$~\cite{Eisenstein:2005su} and observational Hubble data
from~\cite{Riess:2011yx,Stern:2009ep,Gaztanaga:2008xz}. Those data
points with $z>0.9$ in these data sets will be abandoned. As a
result, in our calculations, only 519 SNIa data points, 11 Hubble
data points and one BAO data point are used. In this case, one needs
not to assume the form of $w_{de}(z>0.9)$ and to consider
correlations between $w_{de}(z>0.9)$ and $w_{de}(0\leq z\leq0.9)$.

The cosmological parameters are fitted with the SNIa data~\cite{Su:2011ic} by
 \be
\chi_{SN}^{2}=\sum_{i=1}^{519}\frac{[\mu_{th}(z_{i})-\mu_{ob}^i]^{2}}{\sigma^{2}_i}~,\label{musn}
\ee where the theoretical distance modulus \be
\mu_{th}(z)=5\log_{10}D_L+\mu_{0}~. \ee For a flat
Friedmann-Robertson-Walker universe, the luminosity distance is
\be
D_L=(1+z)\int_{0}^{z}dx/E(x)~, \ee where
 \be
E^2(z)=\Omega_{m0}(1+z)^3+(1-\Omega_{m0})F(z).
  \ee
Here $\Omega_{m0}$
is the current fractional matter density of the universe and function
$F(z)$ depends on the parametrization of $w_{de}(z)$:
\be
F(z)=e^{3\int_0^z\frac{1+w_{de}}{1+x}dx}.
\ee

 One can expand Eq.(\ref{musn}) with respect to $\mu_0$ as \be
\chi_{SN}^{2}=a+2b\mu_{0}+c\mu_{0}^{2} \ee where
\begin{eqnarray}
a & = & \sum_{i=1}^{519}\frac{[\mu_{th}(z_{i};\mu_{0}=0)-\mu_{ob}^i]^{2}}{\sigma^{2}_i},\nonumber\\
b & = & \sum_{i=1}^{519}\frac{\mu_{th}(z_{i};\mu_{0}=0)-\mu_{ob}^i}{\sigma^{2}_i},\label{abc}\\
c & = & \sum_{i=1}^{519}\frac{1}{\sigma^{2}_i}~.\nonumber
\end{eqnarray}
The $\chi_{SN}^{2}$ has a minimum with respect to $\mu_0$,
 \be
\tilde{\chi}_{SN}^{2}=a-{b}^2/{c}~.\label{sn}
 \ee
This way the nuisance parameter $\mu_0$ is reduced, in this work we
will adopt $\tilde{\chi}_{SN}^{2}$ instead of $\chi_{SN}^{2}$.

The BAO distance parameter $A$ is the measurement of BAO peak in the
distribution of SDSS luminous red galaxies~\cite{Eisenstein:2005su}
\be
A=\Omega_{m0}^{1/2}E^{-1/3}(0.35)\left(\frac{1}{0.35}\int_0^{0.35}\frac{dz}{E(z)}\right)^{2/3}.
\ee The value of $A$ is determined to be
$0.469(n_s/0.98)^{-0.35}\pm0.017$, where $n_s=0.963$ is the scalar
spectral index, which has been updated from the WMAP7
data~\cite{Komatsu:2010fb}. The $\chi^2_A$ is defined as \be
\chi^2_{A}=\frac{(A-0.472)^2}{0.017^2}~. \ee

The observational Hubble data can be obtained by using the differential
ages of passively evolving galaxies
 \be
H\simeq-\frac{1}{1+z}\frac{\Delta z}{\Delta t}~. \ee
Here we adopt 11
observational Hubble data points with $z\leq0.9$
from~\cite{Riess:2011yx,Stern:2009ep,Gaztanaga:2008xz}. Those data
points are summarized in Table {\ref{TI}.
\begin{table}
\begin{centering}
\begin{tabular}{|c|c|c|c|c|c|c|c|c|c|c|c|}
\hline
z &0 &0.1&0.17&0.27&0.4&0.48&0.88&0.9&0.24&0.34&0.43\\
\hline
h&0.738 &0.69&0.83&0.77&0.95&0.97&0.9&1.17&0.7969&0.838&0.8645\\
\hline
$\sigma_h$&0.024 &0.12&0.08&0.14&0.17&0.6&0.4&0.23&0.0232&0.0296&0.0327\\
\hline
\end{tabular}
\end{centering}
\caption{11 observational Hubble data with their redshifts from~\cite{Riess:2011yx,Stern:2009ep,Gaztanaga:2008xz}.}
\label{TI}
\end{table}
The $\chi^2_{HUB}$ is defined as: \be
\chi^2_{HUB}=\sum_{i=1}^{11}\frac{[h_{th}(z_i)-h_{ob}(z_i)]^2}{\sigma_{h,i}^2}~,
\ee where $h=H/100\  km\cdot s^{-1}\cdot Mpc^{-1}$.

Finally the total $\chi^2_T$ for three kinds of observational data
is the sum of them: \be \chi^2_T=\tilde
\chi^2_{SN}+\chi^2_A+\chi^2_{HUB}. \ee

\subsection{Methodology and Results}

Now by the piecewise parametrization approaches for $w_{de}$, we get
the constraints of $w_{de}$ from the observational data mentioned in
the above. We will use the 0-spline, liner spline and cubic spline
method, respectively. The best-fitting $w_{de}(z)$, $Q(z)$ and their
$1\sigma$, $2\sigma$ C.L. errors obtained by using three spline
methods are plotted in Fig. \ref{f1}~. One can see that the
constraints of $w_{de}$ from the three methods are consistent with
each other. It is shown that the constraints of $w_{de}$ beyond
$z\sim0.5$ are much weaker than those within $z\sim0.5$, and
$w_{de}=-1$ (the cosmological constant) is still consistent with
present constraints of $w_{de}$ at $2\sigma$ C.L. In addition, there
might exist a rapid transition of $w_{de}$ around $z=0.5$, which is
particularly evident in the fitting results from the 0-spline and
linear-spline methods.

\begin{figure}[h!]
  \includegraphics[width=6in,height=6in]{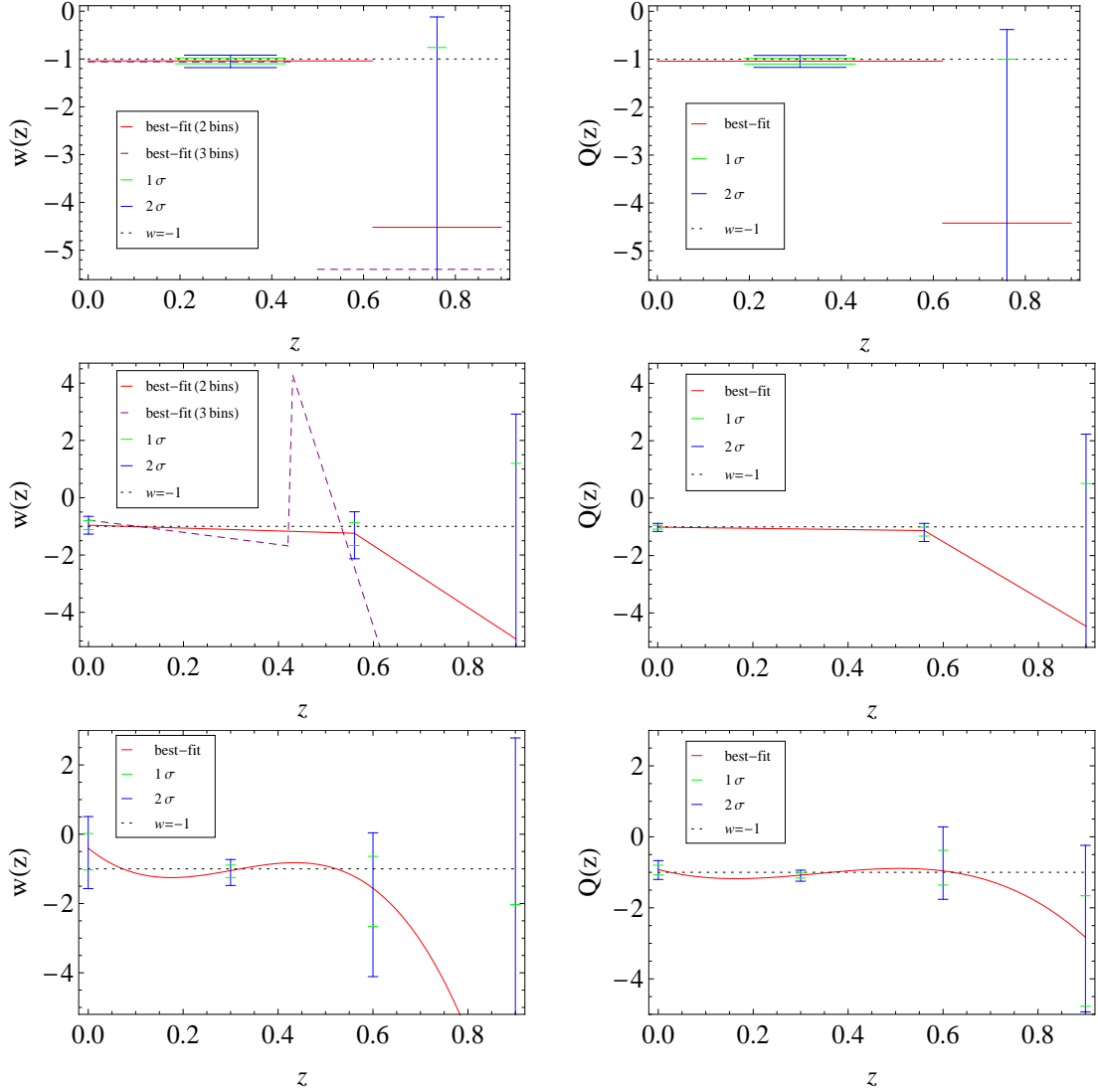}
  \caption{The best fitting $w_{de}(z)$, $Q(z)$ and their $1\sigma$ and $2 \sigma$ C.L. errors
   from the current observational data (SN+A+HUB). The black dotted curves stand for $w=-1$.
  The three figures in left panel are for the correlated $w_{de}(z)$ and the three figures in right panel
  are for the uncorrelated parameters $Q(z)$. The upper two figures  are from the 0-spline method,
    the middle two from the linear-spline method and the bottom two from the cubic-spline method.
  Note that some best fitting results on $w_{de}$ and their errors are not included in figures when they are beyond the range of
  figures. } \label{f1}
\end{figure}

\subsubsection{0-spline method}
In this method, one divides the redshift region under consideration
into $n$ bins, and sets \be w_{de}(z_i<z\leq z_{i+1})=w_i~, \ee
where $w_i$ are constants and $z_i$ are divided positions of bins. Since the number and precision of present
data are still not sufficient enough, at first we divide the
redshift region $[0,0.9]$ into only two bins and treat the divided
position $z_1$ as a free parameter of the model. By fitting the
model with data, we find the best value of divided position $z_1=0.62$. Thus  we
have \be w_{de}(z)=\left\{
\begin{array}{cc}
w_1~,& 0 \le z\leq 0.62 \\
w_2~,&0.62<z\leq0.9
 \end{array}\right..
\ee In the fitting process we have assumed a prior $-20<w_2$.
Otherwise  $w_2$ would go to a very large minus value in MCMC
procedure and thus the downward error of $w_2$ would be extremely large.
Of course the prior will not qualitatively affect the results and
conclusions of this paper. It is found that there is almost no difference
between the errors of $w_i$ and  corresponding $Q_i$. This
shows that the correlation between $w_1$ and $w_2$ is extremely
small in this case.

The fitting results are shown in Table {\ref{TII} and Fig.~\ref{f1}.
It indicates that the constraints of $w_{de}$ are very
good in the whole first bin, since here we have assumed $w_{de}$ to be a
constant and the errors of $w_{de}$ are averaged in each bin.
For the second bin, the errors of $w_2$ are
very large, particularly its downward error, as expected. Of course the fitting
results  depend on the number of bins and the divided manners. Here
the width of the second bin is relatively small and there are only
103 data points in this bin, but the main reason for the weak
constraint of $w_2$ ($w_{de}$ in the second bin) is due to the
high redshift; in the lower redshift region, the same width of bin
and the same number of data points can give much better constraints
of $w_{de}$.

Next we  divide the readshift region $z\in [0,0.9]$ into three bins
to see whether there exist more structures of $w_{de}(z)$ in this
region. Two divided positions are also treated as free parameters,
the positions of them from the best fitting are: $z_1=0.45$ and
$z_2=0.50$. In this case,  the second bin is relatively narrow. The
best-fitting $w_{de}(z)$ is found to be: \ba
w_{de}(z)=\left\{\begin{array}{cc}
-1.06~,&0\leq z\leq0.45\\
5.82~,&0.45<z\leq0.5~~~~.\\
-5.40~,&0.5<z \leq 0.9
\end{array}\right.
\ea This results are plotted in Fig.~\ref{f1},  but the result for
the second bin $w_2$=5.82 is not included because it deviates far
from $w_{de}$ in other bins. The appearance of the narrow second bin
and the large deviation of $w_2$ from the values in other bins might
imply that there is a rapid transition of $w_{de}$ around
$z\sim 0.5$. Mock future data with a rapid transition $w_{de}$
around $z \sim 0.5$ is simulated in the next section and a similar
fitting result from the data is found.

\begin{table}[t!]
\begin{centering}
{\begin{tabular}{c|c|c|c|c}
\hline
h &$\Omega_{m0}$&$w_1\ {\rm and}\ Q_1$&$w_2\ {\rm and}\ Q_2$&$\chi^2_{min}$\\
\hline
\multirow{2}{*}{$0.722^{+0.012+0.025}_{-0.014-0.028}$} &\multirow{2}{*}{$0.275^{+0.022+0.042}_{-0.015-0.033}$}&$-1.04^{+0.06+0.12}_{-0.06-0.14}$&$-4.52^{+3.76+4.40}_{-15.38-15.47}$
&\multirow{2}{*}{499.832}\\
&&$-1.04^{+0.06+0.12}_{-0.06-0.13}$&$-4.42^{+3.42+4.04}_{-15.00-15.05}$\\
\hline
\end{tabular}}
\end{centering}
\caption{The best-fitting values with $1\sigma$ and $2\sigma$ C.L. errors
of parameters from present data in the case with the 0-spline
method.} \label{TII}
\end{table}

\subsubsection{Linear-spline method}

In this case  we set $w_{de}(z)$ as \be w_{de}(z_{i-1}<z\leq
z_i)=w(z_{i-1})+\frac{w(z_i)-w(z_{i-1})}{z_i-z_{i-1}}(z-z_{i-1})~,~~(1\leq
i\leq n) \ee
 where $n$ is the number of bins and $z_0=0$. Still, the
redshift region $z\in[0,0.9]$ is divided into two bins and the best
fitting result gives the divided position to be $z_1=0.56$, which is
very close to the one given by 0-spline method. Here the prior
$-20\leq w(0.9)$ has been assumed. The best fitting values of
the parameters $w(0)$, $w(0.56)$ and $w(0.9)$ and their errors are
shown in Fig.~\ref{f1} and Table III. It is shown that the errors
of $w_{de}$ in the second bin (especially the downward error) increase quickly with
redshift $z$. At $z=0.9$, the constraint of $w_{de}$ becomes
extremely weak.

The best-fitting $w_{de}(z)$ in [0,~0.9] with three bins is also shown in
Fig.~\ref{f1}. The best fitting results for $w_{de}$ are:
$w(0)=-0.78$, $w(0.42)=-1.68$, $w(0.43)=4.28$ and $w(0.9)=-20$. Note
that here we have also assumed the prior $-20\leq w(0.9)$. One
can see from the figure that  there is also a rapid transition of
$w_{de}$ around $z\sim0.5$ in this best fitting, as the case of the
0-spline method.

\begin{table}[h!]
\begin{centering}
{\begin{tabular}{c|c|c|c|c|c} \hline
h &$\Omega_{m0}$&$w(0)\ {\rm and\ }Q(0)$&$w(0.56)\ {\rm and\ }Q(0.56)$&$w(0.9)\ {\rm and}\ Q(0.9)$&$\chi^2_{min}$\\
\hline \multirow{2}{*}{$0.721^{+0.012+0.026}_{-0.013-0.027}$}
&\multirow{2}{*}{$0.276^{+0.026+0.046}_{-0.015-0.035}$}&$-0.96^{+0.16+0.31}_{-0.15-0.31}$&$-1.24^{+0.37+0.75}_{-0.43-0.89}$
&$-4.93^{+6.13+7.85}_{-15.06-15.06}$&\multirow{2}{*}{499.692}\\
&&$-1.01^{+0.06+0.13}_{-0.07-0.15}$&$-1.13^{+0.12+0.25}_{-0.19-0.38}$&$-4.46^{+4.97+6.69}_{-13.02-13.08}$\\
\hline
\end{tabular}}
\end{centering}
\caption{The best-fitting values with $1\sigma$ and $2\sigma$ C.L. errors
of parameters from present data in the case with the linear-spline
method.} \label{TIII}
\end{table}

\subsubsection{Cubic-spline method}
To use the cubic-spline method, we divide $z\in [0,0.9]$ into three
bins with the fixed divided positions as: $z_1=0.3$, $z_2=0.6$. The
results are shown in Fig.~\ref{f1} and Table IV. In this case, once
again, the constraints of $w_{de}$ in the last bin are very weak and
the errors of $w_{de}$ are extremely large. The errors of the uncorrelated parameter $Q(0.9)$
are much smaller than those in other two methods but the errors of
other $Q$'s are larger than those in other two methods, because in
the cubic-spline method $w_{de}$ in different bins are highly
correlated.

\begin{table}[h!]
\begin{centering}
\footnotesize{\begin{tabular}{c|c|c|c|c|c|c}
\hline
h &$\Omega_{m0}$&$w(0)$\ {\rm and}\ $Q(0)$&$w(0.3)$\ {\rm and}\ $Q(0.3)$&$w(0.6)$ \ {\rm and} \ $Q(0.6)$&$w(0.9)$
\ {\rm and} \ $Q(0.9)$&$\chi^2_{min}$\\
\hline
\multirow{2}{*}{$0.720^{+0.013+0.026}_{-0.013-0.027}$} &\multirow{2}{*}{$0.278^{+0.026+0.049}_{-0.015-0.035}$}&$-0.40^{+0.42+0.91}_{-0.63-1.17}$&$-1.05^{+0.16+0.32}_{-0.20-0.43}$
&$-1.56^{+0.92+1.60}_{-1.10-2.55}$&$-9.88^{+7.85+12.66}_{-10.09-10.11}$&\multirow{2}{*}{498.942}\\
&&$-0.92^{+0.12+0.25}_{-0.15-0.28}$
&$-1.08^{+0.06+0.14}_{-0.08-0.17}$&$-0.96^{+0.58+1.24}_{-0.40-0.80}$&$-2.83^{+1.17+2.59}_{-1.94-2.10}$\\
\hline
\end{tabular}}
\end{centering}
\caption{The best-fitting values with $1\sigma$ and $2\sigma$ errors
of parameters from present data in the case with the cubic-spline
method.} \label{TIV}
\end{table}

\section{constraints of $w_{de}$ from future data}

To see the constraint ability on $w_{de}$ from
future observational data, we adopt the characteristics of a
SNAP-like JDEM survey~\cite{Kim:2003mq} to simulate the future SN
data with $0.1<z<1.7$, which include $1998$ SN data points. The redshift distribution
of the mock SN data is shown in Table V, in which 300 supernovae
with $z<0.1$~\cite{sn300,Hojjati:2009ab} are also included. In each redshift bin as
shown in Table V, SN's are assumed to be uniformly distributed.
\begin{table}
\begin{centering}
\begin{tabular}{|c|c|c|c|c|c|c|c|c|c|c|c|c|c|c|c|c|c|}
\hline
z$\rightarrow$ &0.1 &0.2&0.3&0.4&0.5&0.6&0.7&0.8&0.9&1.0&1.1&1.2&1.3&1.4&1.5&1.6&1.7\\
\hline
$N_{bin}$&300 &35&64&95&124&150&171&183&179&170&155&142&130&119&107&94&80\\
\hline
\end{tabular}
\end{centering}
\caption{The redshift distribution of 1998 SN data with $0.1<z<1.7$ for a
SNAP-like JDEM survey~\cite{Kim:2003mq} and 300 SN data with $z<0.1$ from the NSNF~\cite{sn300}.
The redshifts in the first row are the
upper limits of each bin.}
\end{table}

We will use two fiducial models to simulate the mock data:\\
Model I: one assumes a slowly varying equation of state for DE:
 \be
w_{de}(z)=-0.8-\frac{300000}{e^{22/(1+z)}+600000}~; \ee
 Model II:
one has the equation of state with a rapid transition around
$z\sim0.5$ : \be
w_{de}(z)=-0.8-\frac{3\times10^{14}}{e^{100/(1+z)-30}+6\times10^{14}}~;
\ee
In Fig.~\ref{f2} two fiducial $w_{de}(z)$ are plotted.
The form of $w_{de}(z)$ in these two models is also adopted in
\cite{Holsclaw:2010sk}. Both fiducial models have $h=0.72$ and
$\Omega_{m0}=0.28$. Now the distance modulus of $\sim2300$ SN can be
simulated and the corresponding errors are assumed as
\cite{Kim:2003mq,Hojjati:2009ab}: \be
\sigma(z)=\sqrt{\frac{\sigma^2_{obs}}{N_{bin}}+dm^2}, \ee where
$\sigma_{obs}=0.15$, $dm=0.02z/z_{max}$ and $z_{max}$ is the maximum
redshift (here $z_{max}=1.7$). To simulate the effect of other
future observations and alleviate the degeneracy between
$\Omega_{m0}$ and $w_{de}$, we add a prior
$\Omega_{m0}=0.28\pm0.03$. In all three spline methods, the redshift
region $(0,1.7)$ will be divided into three bins. Two divided
positions are still treated as free parameters in the cases of
0-spline and linear-spline methods, while the divided positions of
bins are fixed by hand in the case of cubic-spline method.  All
fitting results are shown in Fig.~\ref{f2}. One can see that three
spline methods give consistent results and the future mock data give
poor constraints of $w_{de}$ beyond $z\sim1$.

\begin{figure}[t]
  \includegraphics[width=6in,height=6in]{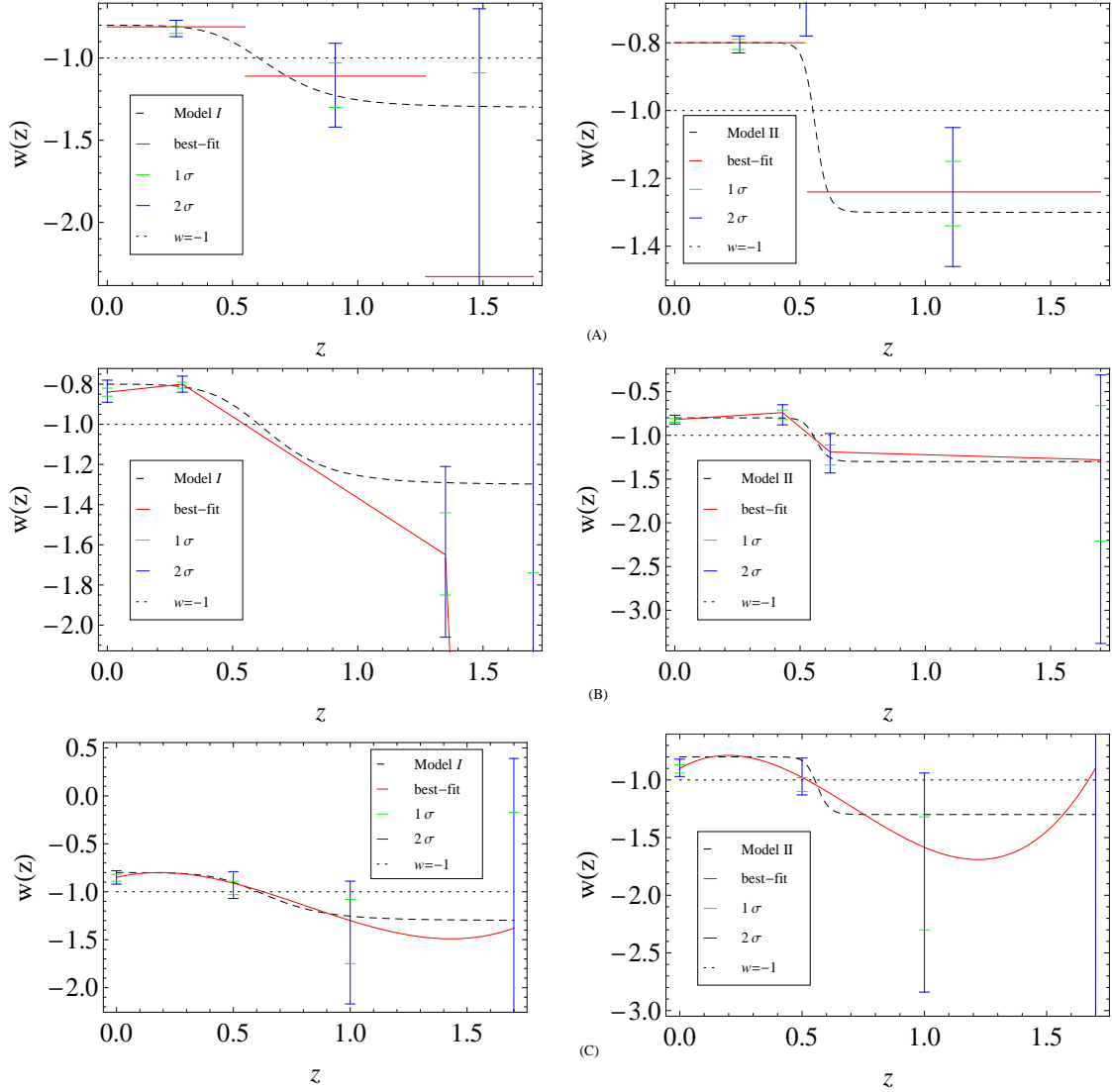}
  \caption{The best fitting $w_{de}(z)$ and their $1\sigma$ and $2\sigma$ C.L. errors from the mock future data.
  The dashed curves are for the fiducial models, and the black dotted curves stand for the cosmological constant
   $w=-1$.
  The three figures in left panel are for model I and the three figures in right panel for model II.
  The upper two figures are from the 0-spline method, the middle ones from the linear-spline method
  and the bottom two figures from the cubic-spline method.} \label{f2}
\end{figure}

\subsection{0-spline method}
For the fiducial model I, the region (0,1.7) is divided into three relatively uniform bins in the best fitting model,
with the divided positions: $z_1=0.55$ and $z_2=1.27$.
The errors of $w_{de}$ increase rapidly with redshift. In the last
bin, the best-fitting $w_{de}$ deviates far from the fiducial model
and the errors of $w_{de}$ are relatively large.

For the rapid transition model II, the best fitting divided
positions are $z_1=0.52$ and $z_2=0.53$. In this case the second bin
is extremely narrow because of the rapid transition of the fiducial
$w_{de}$ around $z\sim0.5$. The best fitting $w_{de}$ in the second
bin is $w_2=0.67$ which deviates from the fiducial $w_{de}$ beyond
$2\sigma$ C.L.~. The errors of $w_{de}$ in the last bin are much
better than those for model I, since here the width of last bin is
much larger than that in model I.

\subsection{Linear-spline method}

In this case the situation is very similar to the case of
the 0-spline method. For model I, the best fitting divided positions of bins are $z_1=0.3$ and $z_2=1.35$.
It is shown that the fiducial $w_{de}$ can be well reconstructed until $z\gtrsim1$.
For model II, the best fitting divided positions are $z_1=0.43$ and $z_2=0.62$.
Here the constraints of $w_{de}$ at high redshift are much better than those for model I, since
the last bin here is much larger than that for model I.
It can be seen that the rapid transition of
$w_{de}$ in model II can be well reconstructed by the linear-spline
method, as shown in Fig.~\ref{f2}, though the width of second bin is
very narrow. The reconstructed $w_{de}$ and its errors here have
finer structures than those in the 0-spline method.

\subsection{Cubic-spline method}

In this case we fix the divided positions of three bins as $z_1=0.5$ and
$z_2=1.0$.  The errors of $w_{de}$ still increase rapidly with
redshift. The fitting results for
model I and model II indicate that this method can reconstruct the
slowly varying  $w_{de}$ well, but it is not good in reconstructing
the equation of state with rapid transition. For model I the errors
of $w_{de}$ are consistent with those from other two methods,
but for model II the errors are much larger than those from other
two spline methods.

\subsection{ Result analysis}

For model I, it is shown that $w_{de}$ can be well reconstructed up to
$z\sim1$. For model II, since  the width of the last bin are always much
larger than that in model I, the errors of $w_{de}$ in the region
beyond $z\sim1$ are much smaller than those in model I (except for
the case of the cubic-spline method, which is not good at describing
a rapid transition $w_{de}$). This means that the fitting results
depend on the divided manner of redshift bins. In our case, we treat
the divided positions as free parameters and then fix their values to the
best-fitting values (in 0-spline and linear-spline methods) or just divide the redshift region uniformly (in the cubic-spline method). In this case, the width of the last bin is
always not large enough to get strong constraints of $w_{de}$ in
that bin. One may use other ways to divide redshift, even setting the last bin large enough by hand. But
a large bin usually will lead to a lose of fine structure of
$w_{de}$ in this case.

In the case of the 0-spline method, $w_{de}$ and its errors are
averaged in each bin, while in the cases with other two spline
methods, the errors of $w_{de}$ increase with  redshift inside each
bin and the reconstructed $w_{de}$ always have finer structures than
that from the 0-spline method. For the cubic-spline method  it is
good at reconstructing the slowly varying $w_{de}$, but not the case
with rapid transitions, and the errors of $w_{de}$ are highly
correlated.

\section{Conclusions}

We have studied the constraint ability on the equation of state
$w_{de}$ of dark energy from the present and simulated future
observational data by piecewise parametrization with the 0-spline,
linear-spline and cubic-spline methods, respectively.
Three spline methods give consistent results of $w_{de}$: 1)
the cosmological constant $w_{de}=-1$ is still consistent with
present data at $2\sigma$ C.L.; 2) current data can constrain
$w_{de}$ well up to $z\sim0.5$ and the future (mock $\sim$2300 SN)
data can constrain $w_{de}$ well  up to $z\sim1$; 3) in high
redshift region, the downward errors of $w_{de}$ are always much larger
than the upper ones; 4) the fitting results from current data by
using the 0-spline and linear-spline methods indicate that there
might exist a rapid transition of $w_{de}$ around $z\sim0.5$.

There are also differences among the fitting results from the three
spline methods. With the 0-spline method $w_{de}$ and its errors get
averaged in each bin, and thus it always gives poor structure of
$w_{de}$. Therefore this method  is suited to be used to confirm
whether $w_{de}$ is a constant (including -1) or not. The
linear-spline and cubic-spline methods give finer structure of
$w_{de}$ than the 0-spline method. The linear-spline method can
reconstruct almost all types of $w_{de}$ in principle, but the
reconstructed $w_{de}$ is always not smooth at the divided positions
of bins, which will lead to deviations of $w_{de}$ from the real
$w_{de}$ around the divided positions. Thus the linear-spline method
is suited to reconstruct non-smooth $w_{de}(z)$, and the positions
where $w_{de}$ suddenly changes can be accurately determined. For
the cubic-spline method, one needs not to search for the
best-fitting divided positions of bins but the redshift region must
be divided into at least 3 bins. It is shown that the cubic-spline
method is not good at reconstructing $w_{de}$ with rapid transitions
and the errors of $w_{de}$ at different bins are highly correlated.
The cubic-spline method is therefore suited to reconstruct a
smoothly varying $w_{de}$.  Basically, a combination of the fitting
results from the three spline methods can reveal the real $w_{de}$.

The fitting results are also affected by divided manners of redshift
bin. Usually a larger width of one bin will lead to a stronger
 constraint of $w_{de}$ there, but fine structure of
$w_{de}$ will be lost. At present, the number and precision of
observational data are still not sufficient to obtain both strong
constraints and fine structures of $w_{de}$. In particular, to
constrain $w_{de}$ at high redshift, a large number of data will be
required~\cite{Weinberg:2012es}.

{\bf Acknowledgements:} This work was supported in part by the
National Natural Science Foundation of China (No.10821504,
No.10975168, No.11035008, No.11147186 and No.11047001), and the
Ministry of Science and Technology of China under grant No.
2010CB833004, and a grant from the Chinese Academy of Sciences and a
grant from Hanzhou Normal University.

\end{document}